\documentclass[preprint,12pt]{elsarticle}

\usepackage{epsfig}

\usepackage{amssymb}

\journal{Astroparticle Physics}

\begin{document}

\begin{frontmatter}

\title{A new contribution to the conventional atmospheric neutrino flux}
\author{Thomas K. Gaisser}

\address{Bartol Research Institute and Dept. of Physics and Astronomy \\
 University of Delaware, Newark, DE, USA}
 
 \ead{gaisser@bartol.udel.edu}
 
 \author{Spencer R. Klein}
 
 \address{Lawrence Berkeley National Laboratory, Berkeley, CA, USA and \\
 University of California, Berkeley, Berkeley, CA, USA}
 
 \ead{srklein@lbl.gov}

\begin{abstract}
Atmospheric neutrinos are an important background to astrophysical neutrino searches, and are also of considerable interest in their own right.  This paper points out 
that the contribution to  
 conventional atmospheric $\nu_e$ of the rare semileptonic decay of $K_S$
becomes significant at high energy.    Although the $K_S\rightarrow \pi e\nu$ branching ratio is very small, the short $K_S$ lifetime leads to a high critical energy, so that, for vertical showers, the inclusion of $K_S$ semileptonic decay increases the conventional $\nu_e$ flux by $\approx 30\%$ at energies above 100 TeV.    In this paper, we present calculations of the flux of $\nu_e$ from $K_S$.  At energies above their critical energies, the  $\nu_e$ fluxes from kaon decay may be simply related to the kaon semileptonic widths; this leads to a near-equality between the flux of $\nu_e$ from $K^+$, $K_L$ and $K_S$.

\end{abstract}


\end{frontmatter}

\section{\label{sec:intro}Introduction}

Atmospheric neutrinos are of interest for understanding cosmic-ray interactions in the atmosphere and as probes of physics, such as neutrino oscillations \cite{Halzen:2008zz,Halzen:2010yj}.  They are also an important background in searches for high-energy astrophysical neutrinos, particularly in searches for a diffuse flux. In diffuse searches, the significance of any signal depends critically on the assumed flux and spectral shape of the atmospheric neutrino background.  

High-energy atmospheric neutrinos are typically divided into two classes: conventional and prompt \cite{Gaisserbook}.  Conventional neutrinos come from the decays of pions and kaons, and the muons produced when pions and kaons decay.   Pions and kaons  have lifetimes long enough so that, at energies above $\approx 1$ TeV, they are likely to interact before they decay; the relative interaction probability increases linearly with energy, so the neutrino spectrum from decays at high energy is softer.  At high energies, where most muons reach the ground before they decay, the principal sources of $\nu_e$ are the decays
$K^+\rightarrow\pi^0 e^+ \nu_e$ and $K_L\rightarrow\pi^- e^+\nu_e (\pi^- e^-\bar{\nu}_e)$.   
It was pointed out recently that $\eta$ and $\eta'$ can decay to $\mu^+\mu^-$ and contribute to the conventional muon flux, but not to the neutrino flux \cite{Illana:2010gh}.   

Prompt neutrinos come from the decays of charmed and bottom hadrons.   These particles decay quickly (critical energies $\sim10^7$~GeV or higher) so the spectral index of neutrinos from their decays is similar to that of the primary cosmic-ray spectrum
in the energy range considered here.  

There have been many calculations of the conventional neutrino flux. Several analytic calculations exist, mostly using the method of Z-moments \cite{Gaisserbook,Lipari:1993hd}.   Other calculations use Monte Carlo simulations, often based on different hadronic interaction models \cite{Fedynitch:2012fs,Honda:2006qj,Barr:2004br}.  

In this work we evaluate the contribution of the rare, semileptonic decay of $K_S$
to the flux of $\nu_e$.  This contribution has been neglected previously because of its low branching ratio, which makes its contribution negligible below $10$~TeV.  For the same reason, this channel is not tracked in CORSIKA~\cite{Heck:1998vt}.  Although the semileptonic branching ratio is very small, the $K_s$ lifetime is very short, so that, in cosmic-ray air showers, it is more likely to decay than to interact.  As a consequence, its contribution to the flux of $\nu_e$ is one power harder in energy than those
from $K_L$ and $K^\pm$, so that at sufficiently high energy it contributes a significant
fraction of the total.

\section{Electron neutrinos from $K_S$}

The characteristic energy $\epsilon_i$ that characterizes whether an unstable
particle is more likely to interact or decay in the atmosphere is
\begin{equation}
\label{eq:Ecrit}
\epsilon_i = \frac{m_i c^2 h_0}{c\tau_i},
\end{equation}
where $m_i$ and $\tau_i$ are the particles mass and lifetime, and $h_0$ is a scale height in the atmosphere, typically 6400 m \cite{Gaisserbook}.   The energy at which hadronic interactions become important depends on the atmospheric density, which varies with zenith angle $\theta_z$.  Interactions become predominant at energies above the critical energy:
\begin{equation}
E_{\rm crit} = \epsilon/\cos(\theta_z)
\end{equation}
If the particle energy is higher than $E_{\rm crit}$, then it is likely to interact before it can decay.  
Below the critical energy for a given channel, the spectrum of neutrinos from kaon decays 
closely matches that of the cosmic-ray spectrum, roughly $dN/dE\approx E^{-2.7}$, with the neutrino taking an average of roughly 25\% of the kaon energy for $K\rightarrow\pi e\nu$ decays.  At energies above the critical energy, the increasing interaction probability softens the spectrum by $E^{-1}$, to $dN/dE\approx E^{-3.7}$.  Table 1 shows the semi-electronic branching ratios and critical energies for different types of kaons, along with those of charmed hadrons for comparison. 

\begin{table}[t]
 \begin{tabular}{|l|c|c|c|c|}
 \hline
Type & Mass & Br($K\!\rightarrow\!\pi e \nu)$ & Lifetime & Characteristic energy \\
		  & (MeV)  & (\%) & (s) & (GeV) \\
\hline
$K^+$   &  493.6 & 5.04   & $1.24\times10^{-8}$    & 850 \\
$K^0_L$ &  497.6 & 40.55 &  $5.12\times10^{-8}$    & 210 \\
$K^0_S$ &  497.6 & 0.07   & $0.90\times10^{-10}$  & 120,000 \\
Charm    &  $\approx 1800$ &  & $\approx 10^{-12}$ & $\approx 4\times10^{7}$ \\
 \hline
 \end{tabular}
\caption{Masses, semi-electronic branching ratios, lifetimes and characteristic energy for
the different kaon types, and, for comparison, charmed hadrons \cite{Beringer:1900zz,Gaisser:2013ira}. 
\label{tab:kaon}}
\end{table}

The $\nu_e$ flux from $K_S$ decay may be easily determined by reference to the $\nu_e$ flux from
$K_L$ decays.  $K_S$ and $K_L$ are produced at the same rate in air showers, and the $K\rightarrow\pi e\nu_e$
kinematics are almost identical.    At low energies, the $K_S$ contribution to the atmospheric $\nu_e$ flux is small, 
reduced by the ratio of the branching ratios $K_S/K_L$: $0.07/40.55 = 0.0017$.
At higher energies, above 210 GeV$/\cos(\theta_z)$, the spectrum of $\nu_e$ from $K_L$-decay softens to $E^{-3.7}$,
while the spectrum of $\nu_e$ from $K_S$ remains unchanged.  Thus, the relative $\nu_e$ contribution increases linearly with the 
energy.    At the $K_S$ critical energy of 120 TeV$/\cos(\theta_z)$, the ratio has increased by
$\epsilon_{K_S}/\epsilon_{K_L}\approx 588$, and the $K_S$ and $K_L$ contributions to the
$\nu_e$ flux are equal!  This is not just a fortuitous numerical coincidence.  It happens
because the lifetime is inversely related to the total decay width, and the branching ratio is
the ratio of the semileptonic width to the total width.  With $\epsilon\propto 1/\tau=\Gamma_{tot}$ and 
$Br(K\rightarrow\pi e\nu) = \Gamma_{sl}/\Gamma_{tot}$,
as long as the $K_S$ and $K_L$ are produced in equal numbers, the ratio of the $\nu_e$ fluxes for neutrino energies above the two $E_{\rm crit}$ is
\begin{eqnarray}
\hfill \frac{\phi(\nu_e {\rm from} K_S)}{\phi(\nu_e {\rm from} K_L)} 
&=& 
\frac{Br(K_S\rightarrow\pi e \nu)}{Br(K_L\rightarrow\pi e \nu)}
\frac{\epsilon_{K_S}}{\epsilon_{K_L}}\hfill  \\ \nonumber
& = &
\frac{\Gamma_{SL}(K_S)/\Gamma_{Tot}(K_S)}{\Gamma_{SL}(K_L)/\Gamma_{Tot}(K_L)} 
\frac{(1/\tau_{K_S})}{(1/\tau_{K_L})} 
=  1.
\end{eqnarray}

A similar argument applies for $K^+\rightarrow\pi^+ e\nu_e$, which has a similar mass and semileptonic width as the $K_L$ and $K_S$.  However, associated production in reactions like $pp=\rightarrow K^+\Lambda p$ is different for $K^+$ than for the $K^0$ and $\bar{K}^0$ from which the $K_L$ originate.  As a consequence, the contribution of charged kaons to the flux of $\nu_e$ is not exactly equal to that of $K_L$.

Neglecting for the moment associated production, at energies $E_\nu > E_{\rm crit}$, the inclusion of $K_s$ increases the $\nu_e$ flux by about 50\%.   For quasi-vertical angles of incidence, this increase occurs at energies of $\approx 100$ TeV, which is the range in which most current searches for extra-terrestrial neutrinos are focused.  At higher energies, the enhancement is large for a wider angular range, but the conventional $\nu_e$  flux is overshadowed by the prompt flux.  

A similar enhancement occurs for $\nu_\mu$, via $K_S\rightarrow \pi\mu\nu_\mu$.  However, because of the large $\nu_\mu$ contribution from two-body decays of charged kaons and pions, it is much less significant. 
Figure 7 of~\cite{Illana:2010gh} gives the relative contribution to $\nu_\mu$ production of $\pi^+$, $K^+$, $K_L$ and $\mu$ decay.  $K^+$ decay dominates at energies above 500 GeV; the contribution from $K_L$ is negligible, so, at higher energies, the $K_S$ contribution will remain small.  

There are additional $\nu_e$ contributions from the semileptonic decays of strange baryons like the $\Lambda$ and $\Sigma$; some of these baryons have semileptonic branching ratios similar to that of the $K_S$.  However, their production rates are lower, and the neutrino carries only a relatively small fraction of the incident baryon momentum.  So, their contribution to the total flux should be small.  

\section{Flux calculations}
We extend the flux calculation described in Ref.~\cite{Gaisser:2013ira} to include the contribution
of $K_S\rightarrow\pi e\nu_e$.  The calculation is a generalization of the scaling
solutions of the coupled cascade equations for hadronic cascades in the
atmosphere~\cite{Gaisserbook} in which the spectrum weighted moments are
allowed to depend on energy in order to take account of the non-power-law
behavior of the primary spectrum (the knee).  The Z-factors for production
of charged kaons, for example, are generalized to 
\begin{equation}
Z_{NK^\pm}(E)\,=\,\int_E^\infty\,{\rm d}E'\frac{\phi_N(E')}{\phi_N(E)}
\frac{\lambda_N(E)}{\lambda_N(E')}\frac{{\rm d}n_{K^\pm}(E',E)}{{\rm d}E}.
\label{eq:TIG-Z}
\end{equation}
Here $\lambda_N(E)$ is the nucleon interaction length, d$n_{K^\pm}$ is the
number of charged kaons produced in d$E$ by nucleons of energy $E'$, and $\phi_N(E)$
is the spectrum of nucleons.  This method was proposed in Ref.~\cite{Gondolo:1995fq},
and is a good approximation if the energy dependences are smooth.  Simple forms
for the hadronic cross sections~\cite{Gaisser:2001sd} are used to interpolate and extrapolate tabulated values~\cite{Gaisserbook}
of the spectrum weighted moments.  For the calculation of the neutrino fluxes the
spectrum of nucleons per GeV/nucleon is needed, assuming validity of the superposition
approximation in which bound nucleons produce mesons as if they were free.  We
use the spectrum of nucleons from Model H3a of Ref.~\cite{Gaisser:2012zz}.

The basic equation  
for the flux of $\nu_e+\bar{\nu}_e$ at
sufficiently high energy so that the contribution from muon decay can be neglected 
($>\sim 1$~TeV/$\cos\theta$) is
\begin{eqnarray}
\phi_\nu(E_\nu)\, =  \,\phi_N(E_\nu) &\times &
 \left\{{Z_3\,b_{K^+e3}(Z_{NK^+}+Z_{NK^-})\over 1 + 
B_3\cos\theta\, E_\nu / \epsilon_K}\right. \nonumber \\
& & \left.+\,\,\,{Z_{3}\,b_{K_Le3}Z_{NK_L}\over 1+B_3^*\cos\theta\, E_\nu / \epsilon_{K_L}}\right.\nonumber \\
& & \left. +\,\,\,{Z_3\,b_{K_Se3}Z_{NK_S}\over 1+B_3\cos\theta\, E_\nu / \epsilon_{K_S}}\right\}.
\label{eq:angular}
\end{eqnarray}
Here $Z_3\approx 0.134$~\cite{Lipari:1993hd} is the spectrum-weighted moment for the $K_{e3}$ decay at low energy (when $E_\nu\ll \epsilon_{K_x}$).  The branching ratios $b_{K_xe3}$ are for each kaon 
flavor to the $K_{e3}$ mode, and $Z_{NK_x}$ is the spectrum weighted moment
for a nucleon to produce a kaon of type $x$.  The denominator interpolates
between the low and high-energy behavior, where low and high are defined
relative to $\epsilon_{K_x}$ for each neutrino source.  Explicitly,
\begin{equation}
B_3\approx \frac{0.134}{0.061}\left(\frac{\Lambda_K-\Lambda_N}{\Lambda_K\ln\frac{\Lambda_K}{\Lambda_N}}\right) = \frac{Z_3}{Z_3^*}\left(\frac{\Lambda_K-\Lambda_N}{\Lambda_K\ln\frac{\Lambda_K}{\Lambda_N}}\right)
\label{eq:atten}
\end{equation}
where $Z_3^*\approx 0.061$~\cite{Lipari:1993hd} is the high energy value of the spectrum weighted
moment for $K_{e3}$ when the factor $\epsilon_{K_x}/E$ weights the decay
by an extra power of $1/E$.  $Z_3$ and $Z_3^*$ account for the
fraction of the kaon momentum carried by the neutrino.
The branching ratios and critical energies
are listed in Table~\ref{tab:kaon}. 
The factor 
$$\left(\frac{\Lambda_K\ln
\frac{\Lambda_K}{\Lambda_N}}{\Lambda_K-\Lambda_N}\right)  $$
in Eq.~\ref{eq:atten} arises from the integral over atmospheric depth of the
kaon spectrum multiplied by the probability of meson decay in the high-energy limit,
\begin{equation}
\sim\int\frac{{\rm d}X}{\lambda_K}\frac{\epsilon_{Kx}}{E_\nu\cos\theta X}\left(e^{-X/\Lambda_K}-e^{-X/\Lambda_N}\right).
\end{equation}
 
 \begin{figure}[thb]
\begin{center}
\epsfig{file=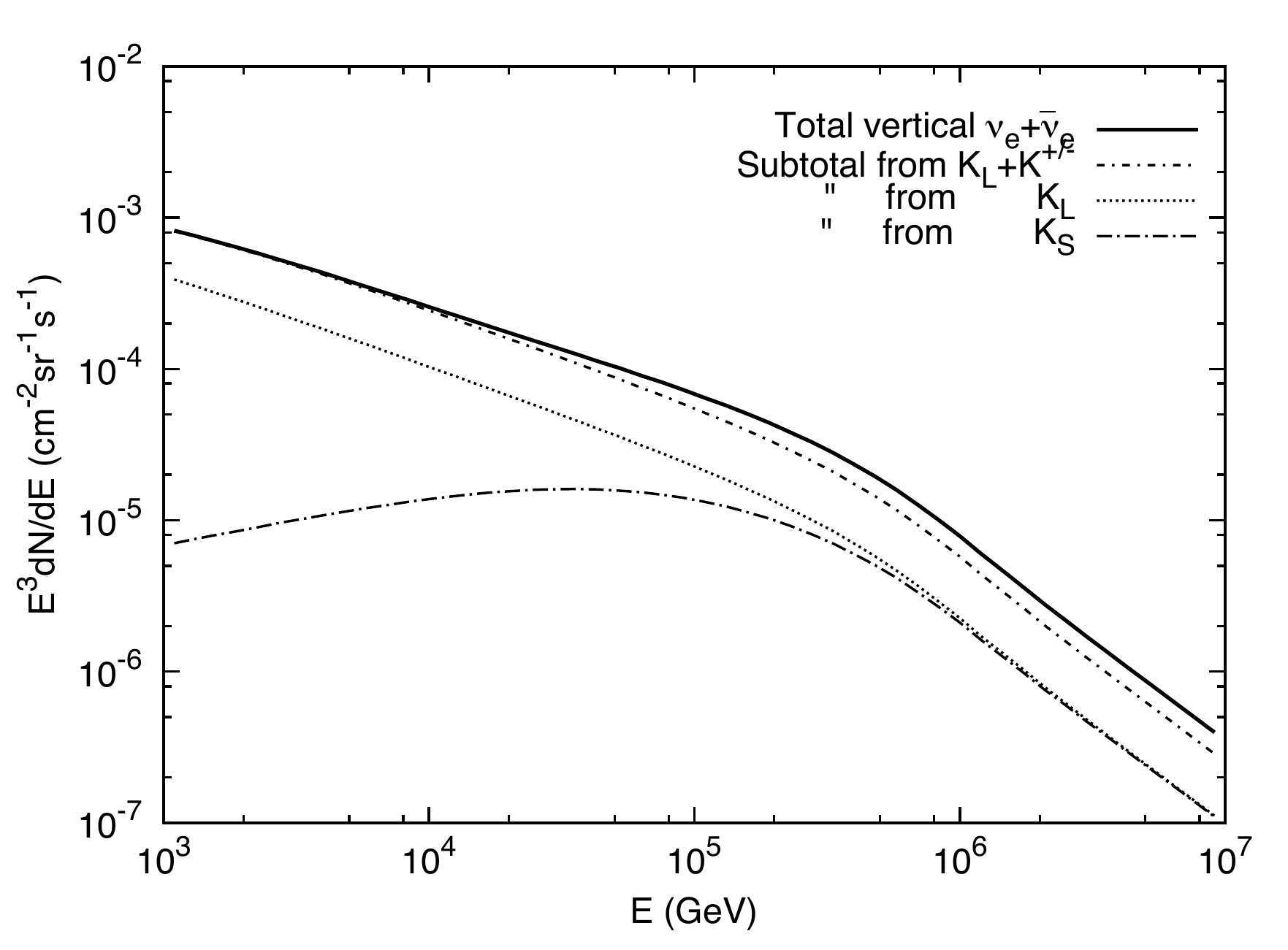,width=2.7in}\epsfig{file=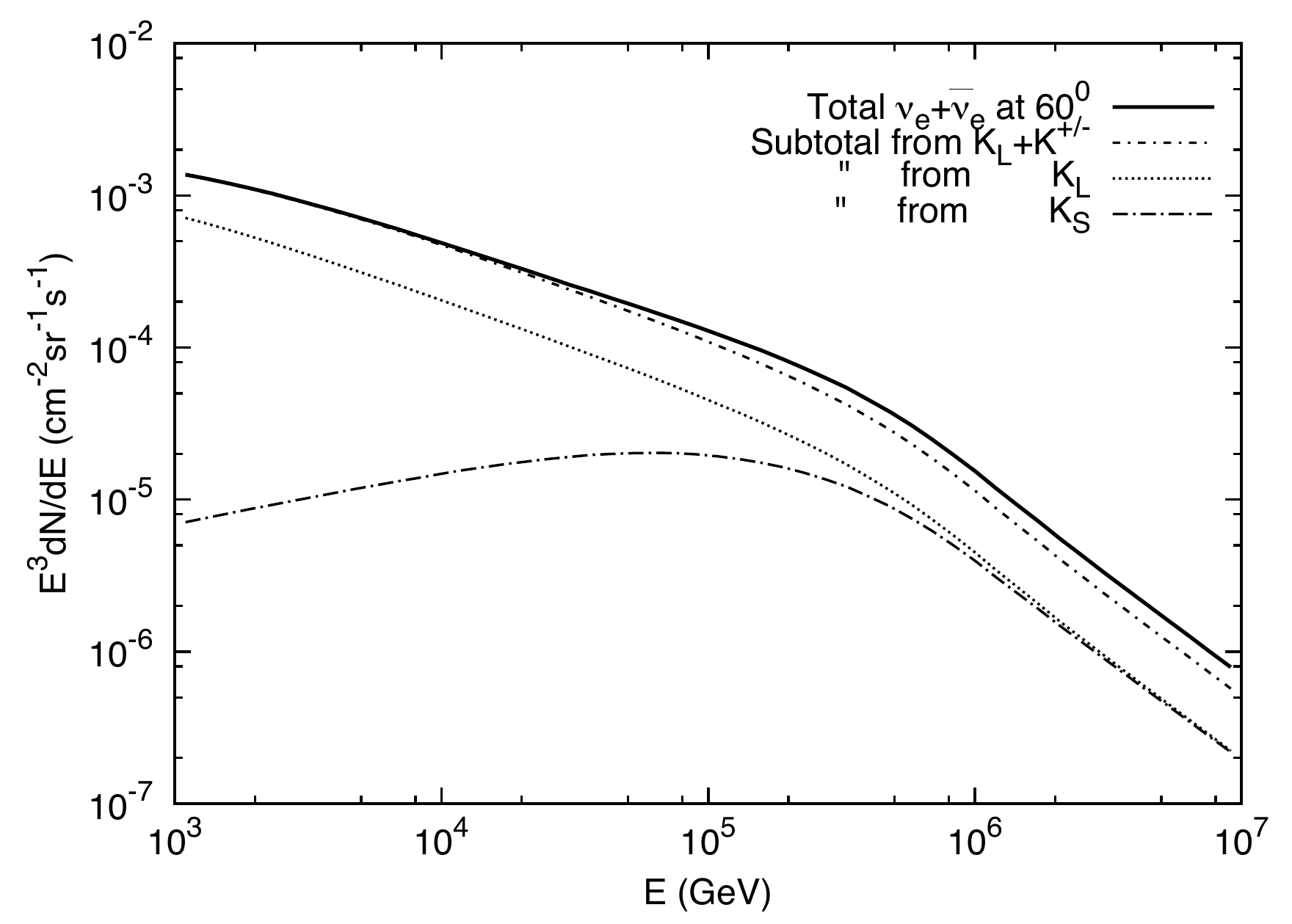,width=2.7in}
\caption{Electron neutrino flux showing the contribution of $K_S$ seraprately.
Left: vertical; Right: $60^\circ$.}
\label{fig:E3dndE}
\end{center}
\end{figure}

\begin{figure}[thb]
\begin{center}
\epsfig{file=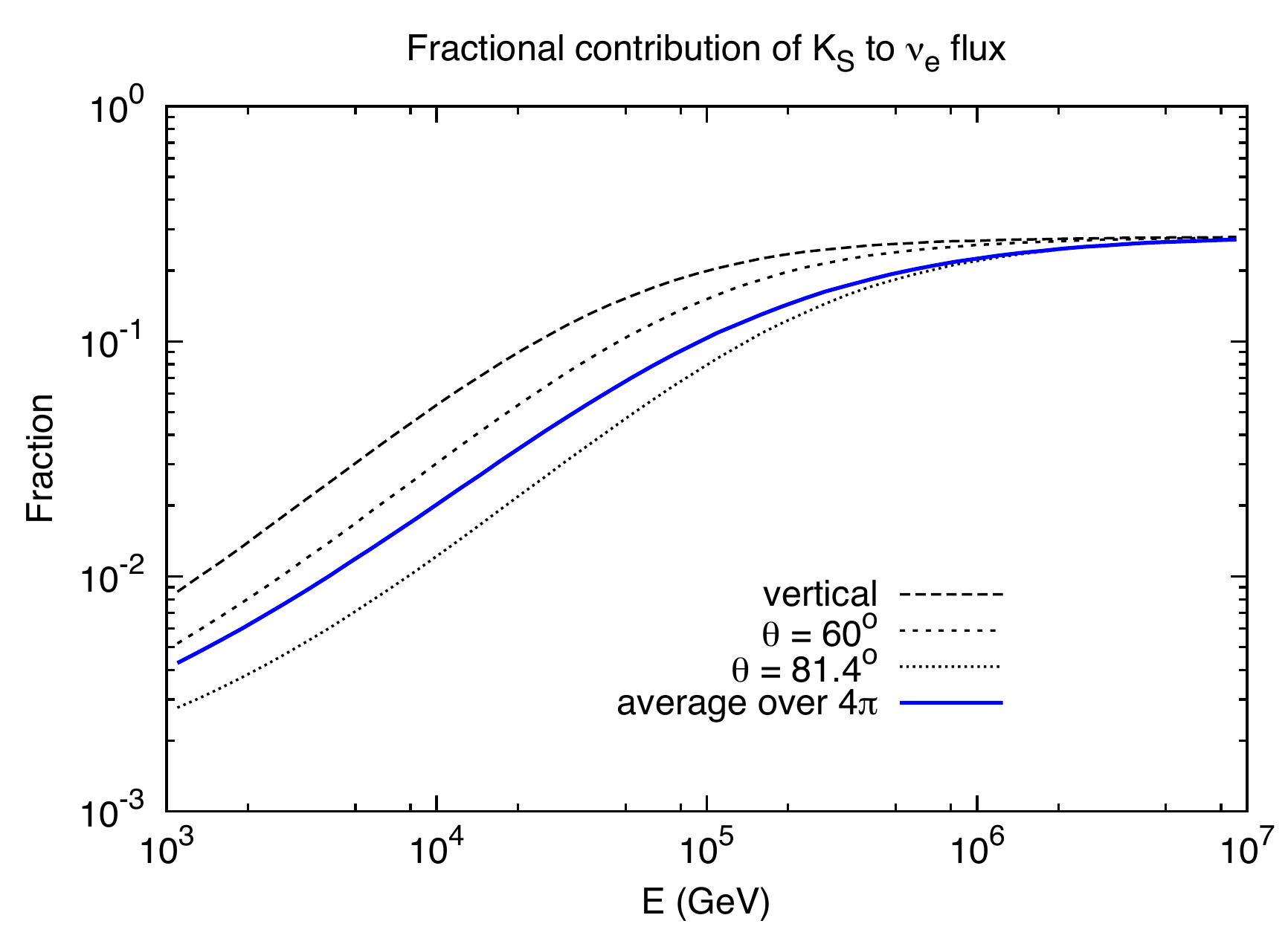,width = 4in}
\caption{Fractional contribution of $K_S$ to the flux of $\nu_e$ at several zenith angles
and averaged over all directions.}
\label{fig:fraction}
\end{center}
\end{figure}

We use a set of Z-factors described in the Appendix to evaluate the the fluxes
of electron neutrinos from Eq.~\ref{eq:angular}.  Results for the flux of ($\nu_e+\bar{\nu}_e$) at two zenith angles are shown in Fig.~\ref{fig:E3dndE}.
Figure~\ref{fig:fraction} illustrates how the fractional contribution of
$K_S$ evolves with energy at different zenith angles.  The critical
energy for $K_S$ is $\approx 120$~TeV$/\cos\theta$, and the neutrino flux from
$K^\pm$ and $K_L$ increases significantly toward the horizontal.  As a consequence,
 when the flux is averaged over zenith angle, the onset of the saturation of the $K_S$
contribution is delayed by an order of magnitude to $E_\nu\sim 1$ PeV.

The $K_L$ and $K_S$ states are not directly produced in hadronic interactions.
Instead they are rotations of $K^0=(d\bar{s})$ and $\bar{K}^0=(\bar{d}s)$.  
Between production and decay, the $K^0=(d\bar{s})$ and $\bar{K}^0=(\bar{d}s)$ states mix through the mass eigenstates $K_S$ and $K_L$. The probability for a neutral kaon to interact or decay as $K^0\rightarrow \pi^0 e^+\nu_e$ or $\bar{K}^0\rightarrow\pi^0 e^-\bar{\nu}_e$ at a later time is given by the probability to find the system in one state or the other. When the states are fully mixed, equal numbers of $\nu_e$ and $\bar{\nu}_e$ are produced
in the Ke3 decays of $K_L$.
In the high energy limit, however, when the neutral kaons interact before they are fully evolved, then asymptotically the relative number of $\nu_e$ and $\bar{\nu}_e$ reflects the
relative importance of $K^0\rightarrow \nu_e$ and $\bar{K}^0\rightarrow\bar{\nu}_e$ at production just as the
neutrinos from decay of charged kaons reflect the relative production of $K^+\rightarrow\nu_e$ and $K^-\rightarrow\bar{\nu}_e$.  Regeneration of $K_S$
when neutral kaons interact in the atmosphere may also affect the
flux of neutrinos.

\begin{figure}[htb]
\begin{center}
\epsfig{file=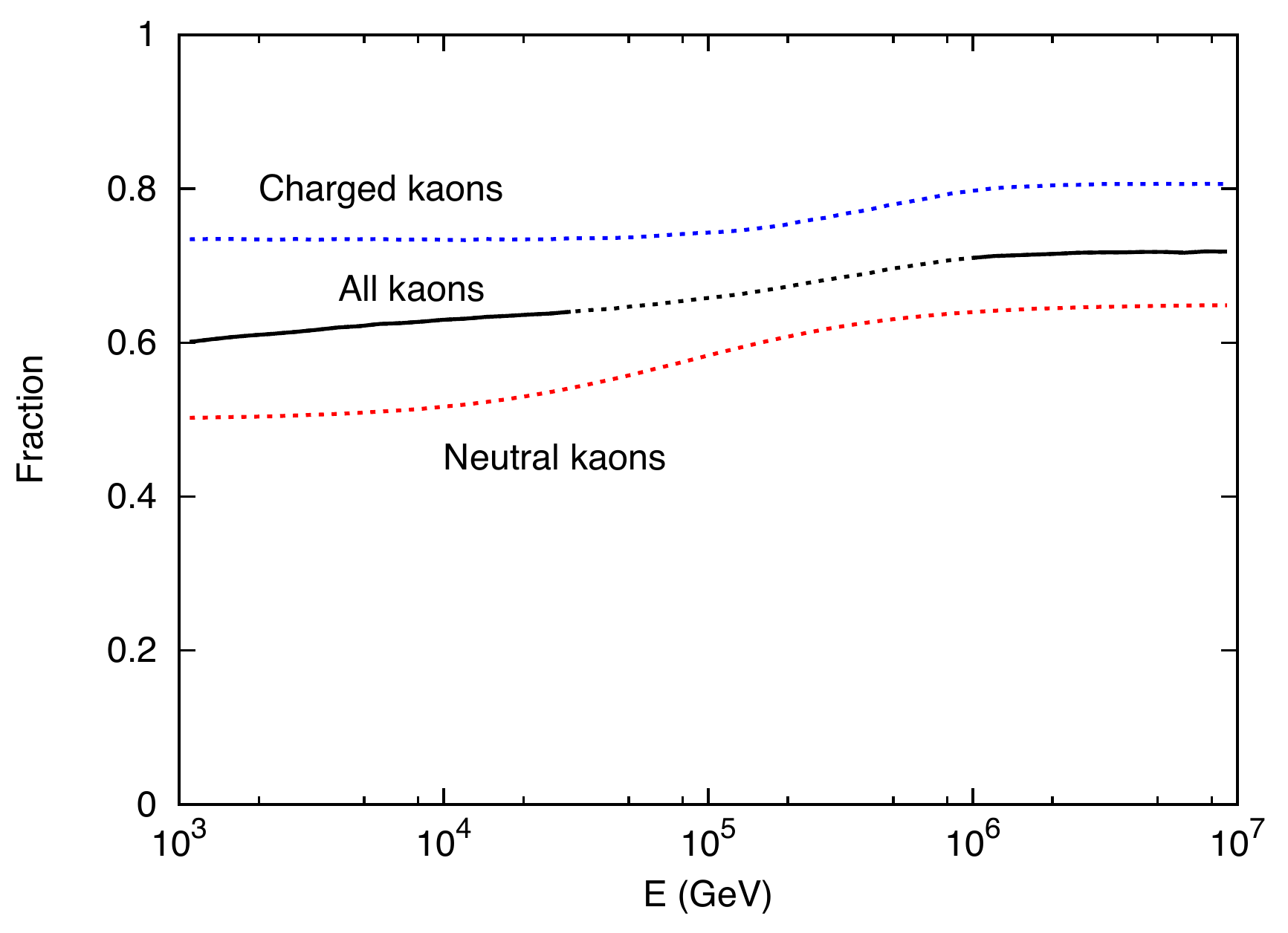,width = 4in}
\caption{Fraction of atmospheric $\nu_e$ in the total flux of $\nu_e+\bar{\nu}_e$.  The fractions are shown separately for
charged kaons (top/blue), neutral kaons (bottom/red) and all kaons.
}
\label{fig:oscillation}
\end{center}
\end{figure}

The effects of kaon oscillations in the cascade equations from which Eq.~\ref{eq:angular} follows
can be approximately accounted for by reference to the standard treatment 
of kaon oscillation in Ref.~\cite{Perkinsbook}.  The oscillation frequency is determined
by the $K_L$--$K_S$ mass difference 
 $\Delta=m_{K_L}-m_{K_S}=0.592\times 10^{10}$~s$^{-1}$~\cite{Beringer:1900zz}.  Thus the oscillation period is $t^*\approx 1.7\times 10^{-10}$~s.  Whether the neutral kaons 
are more likely to interact before or after the oscillations are fully evolved depends
on the structure of the atmosphere in the same way as the competition between decay
and interaction.  As a consequence, the transition between the two regimes is given by a formula like Eq.~\ref{eq:Ecrit} with $\tau_i$ replaced by $t^*\approx 2.1\tau_S$.  The characteristic $\epsilon_{\rm osc}\approx 56$~TeV is fortuitously close to the critical energy for $K_S$ of $120$~TeV.  At higher energies the decays reflect the 
excess of $K^0$ over $\bar{K}^0$ at production.

For electron neutrinos from
Ke3 decays of charged kaons we expect approximately 74\% $\nu_e$ and 26\% $\bar{\nu}_e$ at all energies.  This estimate follows from the values of the respective Z-factors and the relative abundance of protons and neutrons in the primary spectrum of nucleons (see Appendix, Eq.~\ref{eq:Kch2nue}).
Decays of $K_L$ give equal numbers of neutrinos and antineutrinos at low energy.
For $E_\nu > \epsilon_{\rm osc}\approx 50$~TeV, however, the ratio increases, because the decays of neutral kaons still reflect the excess of $K^0$ over
$\bar{K}^0$ at production.  From the numerical approximations of $Z_{K^0}$ and
$Z_{\bar{K}^0}$ in the Appendix, we estimate that the
asymptotic ratio is approximately 65\% to 35\% in favor of $\nu_e$.  Figure~\ref{fig:oscillation} shows how the fraction $\nu_e/(\nu_e+\bar{\nu}_e)$ evolves with energy.  The increase for charged kaons comes from the forward, associated production
of $K^+$, which is less affected by the steepening of the spectrum than the $K^-$.
The increase for neutral kaons comes mainly from the $K^0$--$\bar{K}^0$ asymmetry.
The $\nu_e/\bar{\nu}_e$ ratio is significantly greater than one and increases with energy.
This fact needs to be accounted for in evaluating the atmospheric neutrino background
at all energies.

Several approximations are implicit in Eq.~\ref{eq:angular}, which is a
solution of the atmospheric cascade equations in which production of
pions and kaons from nucleons are accounted for, but production of kaons
by pions is neglected.  In the context of a calculation
of the inclusive rate of neutrinos from a steep primary spectrum, 
such a contribution is proportional to a product of
two small Z-factors and therefore very small.  
A potentially more important approximation
in the present context is the neglect of transitions from one type
of kaon to another.  Accounting for the cross terms would require
solving a set of matrix equations which would be a generalization
of the approach of Ref.~\cite{Lipari:1993hd} to include $K_S$
as well as $K_L$ via production of $K^0$ and $\bar{K}^0$. 
We have checked the solutions of Ref.~\cite{Lipari:1993hd} that track separately the charge-exchange
 process $K^+\leftrightarrow K_L$
to insure that the simpler approximation of Eq.~\ref{eq:angular} is numerically accurate.
We also checked that neglect of the $N\rightarrow\pi^\pm\rightarrow K$ channel
leads to changes at the 1\% level.  
 
The production of kaons by kaons is included in Eq.~\ref{eq:angular} through
the attenuation lengths for each channel defined by
\begin{equation}
\label{K2K}
\Lambda_K\,=\,\frac{\lambda_K}{1-Z_{KK}},
\end{equation}
where $\lambda_K$ is the kaon interaction length.  
The combination in which the attenuation lengths enter the
solution of the cascade equations (Eq.~\ref{eq:atten}) is
a quantity of order one  and does not change 
much over the range of physically possible values of the attenuation lengths.
The asterisk on $B_3^*$ in the term of Eq.~\ref{eq:angular} for $K_L$ allows for
the possibility of suppression of the attenuation length for $K_L$ due to loss to
$K_S\rightarrow 2\pi$ as a consequence of kaon regeneration.  For example, reducing $Z_{KK}$ by the maximum amount possible ($1/2$) in the attenuation length for $\Lambda_{K_L}=\lambda_K/(1-Z_{KK})$ reduces the
contribution of $K_L$ to $\nu_e+\bar{\nu}_e$ by no more than $6$\%.

\begin{figure}[thb]
\begin{center}
\epsfig{file=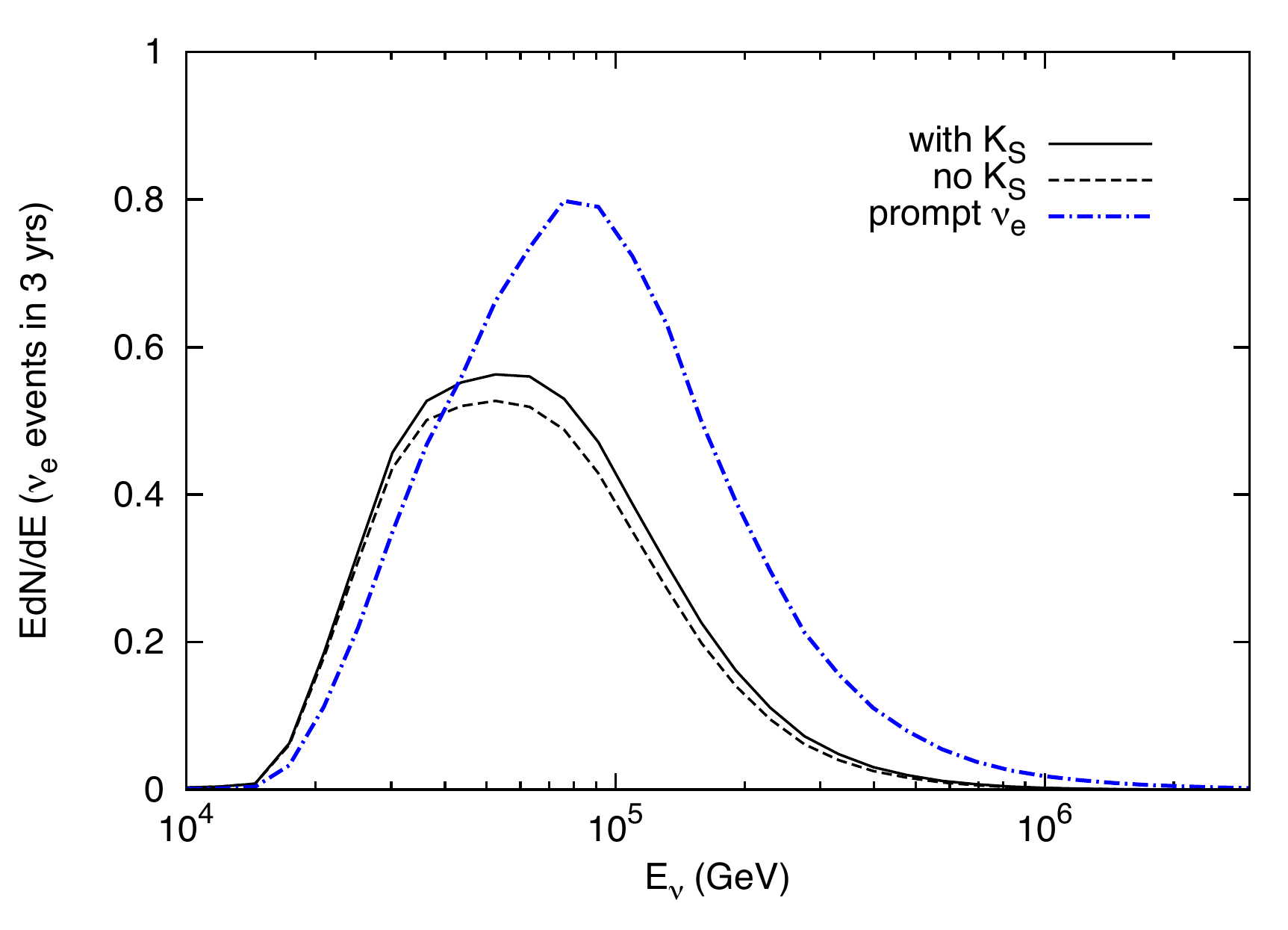,width = 4in}
\caption{Differential distribution of events induced by atmospheric $\nu_e$ in three years of IceCube data with the acceptance as defined in Ref.~\cite{Aartsen:2013jdh}.  The
``prompt'' contribution is for the charm production model of Ref.~\cite{Enberg:2008te},
modified to take account of the knee in the primary cosmic-ray spectrum.}
\label{fig:compare}
\end{center}
\end{figure}

\section{Implications}

Previous measurements of the $\nu_e$ flux have focused on the energy range below a few TeV \cite{Aartsen:2012uu,Daum:1994bf,Aartsen:2013vca,Abbasi:2011ui}, so have not been sensitive to the $K_S$-induced component.  However, searches for extra-terrestrial neutrinos that are sensitive to $\nu_e$ have focused recently on considerably higher energies \cite{Aartsen:2013bka,Aartsen:2013jdh,Aartsen:2014gkd,ANTARES}, and so could be sensitive to the $K_S$ induced component.  The signature is an increase in the atmospheric
$\nu_e$  background with a zenith-angle-dependent inflection point (steepening of the slope) at roughly $30$~TeV$/\cos(\theta_Z)$ (see Fig.~\ref{fig:E3dndE}).   At energies below or near $\epsilon_{K_S}=120$~TeV,
this contribution is nearly isotropic, matching the angular distribution of a prompt atmospheric or diffuse astrophysical flux.  In this respect, the angular distribution is similar to that of prompt neutrinos.   

In order to quantify this effect, we have used the effective areas published in the
supplemental material of Ref.~\cite{Aartsen:2013jdh} and folded them with 
the electron neutrino
fluxes with and without the $K_S$ contribution.  The result is shown in Fig.~\ref{fig:compare} integrated over the three-year live time of Ref.~\cite{Aartsen:2014gkd}.  The effective areas include absorption in the Earth for
neutrinos from below the horizon.  We have also included the effect of the 
neutrino self-veto for events from the Southern sky as calculated in~\cite{Gaisser:2014bja}.  Adding the contribution from $K_S$ increases the
atmospheric $\nu_e$ background by 8\% (from 0.96 to 1.04 events in 3 years).  
When the $\nu_e$ excess is accounted for in the rate calculation, there
is a further increase of 1\% arising from the fact that in the $100$~TeV 
range, the charged current neutrino cross section is still $\sim 15$\%
larger than for anti-neutrinos.  To put the $K_S$ contribution in context,
in the model of Ref.~\cite{Enberg:2008te} corrected for the knee in the primary spectrum, the predicted number of $\nu_e$ from charm decay to the atmospheric
background in the same time interval is 1.48 events. 

In the energy range 10-100 TeV, the relative enhancement of the background 
from the $K_S$ channel is largest in the near-vertical direction, where the conventional atmospheric $\nu_e+\bar{\nu}_e$ flux is the smallest.  Between that and the self-veto~\cite{Gaisser:2014bja}, 
near-vertical neutrinos have much higher signal (astrophysical $\nu$) to background (atmospheric $\nu$) than near the horizon.   In this region, $\nu$ from $K_S$ are an important contributor to event-by-event background estimates. 
 
\section{Conclusions}

We have identified a hitherto overlooked contribution to the conventional atmospheric $\nu_e$ flux, from
$K_S\rightarrow\pi e \nu$. It is potentially significant at energies above 10 TeV, and,  asymptotically, it is nearly equal in magnitude to the components from $K_L$ and $K^+$ decays.  The equality between $K_L$ and $K_S$ components at high energies is independent of the hadronic interaction model that is used to estimate their flux, as long as the two are produced at equal rates.  Using a numerical solution of the cascade equations, we
have evaluated the magnitude of the $K_S$ contribution and find that in practice it
makes a small ($\sim10$\%) increase in an already small background in recent IceCube
analyses aimed at astrophysical neutrinos with energies in the $100$~TeV range and above.

ACKNOWLEDGMENTS.  We are grateful to David Seckel for calling our attention to
the importance of kaon oscillations and regeneration in the context of this paper
and for his suggestions for the text.  Work on this paper began while one of us (TKG)
was participating in a program of the 
Munich Institute for Astro- and Particle Physics (MIAPP) of the DFG 
cluster of excellence ``Origin and Structure of the Universe".
This work was supported in part by U.S. National Science Foundation under grants 
PHY-1307472 and PHY-1205809 and the U.S. Department of Energy under contract numbers DE-AC-76SF00098 and DE-FG02-12ER41808.  

\section*{APPENDIX: Numerical approximations for Z-factors}
We relate the production Z-factors for neutral kaons
in Eq.~\ref{eq:angular} to the Z-factors for production of charged kaons
by assuming that kaon production by nucleons consists of two components:\\
(a) associated production in which a valence di-quark in the projectile nucleon
picks up an ${s}$ quark to produce a forward hyperon ($\Lambda$ or $\Sigma$ of the appropriate charge) and \\(b) production of
${K}$-$\bar{K}$ pairs.  A $K^-$ cannot be produced in association with a single hyperon, so we make the approximation that $K^-$ is always produced as a member of a kaon pair.   
 As a starting point for calculation of the energy-dependent Z-factors by Eq.~\ref{eq:TIG-Z} we use the numerical values of $Z_{pK^+} = 0.0090$ and $Z_{pK^-}= 0.0028$ tabulated in~\cite{Gaisserbook} for a differential spectral index of $2.7$.
We denote the production Z-factor for associated production by a proton as
\begin{equation}
Z_A=Z_{pK^+}-Z_{pK^-} \approx 0.0062.
\label{eq:associated}
\end{equation}

Writing each channel as a sum of associated production and $K^+$-$K^-$ pair production
 leads to
\begin{eqnarray}\label{eq:ZK}
Z_{pK^+} & = & Z_A +Z_{pK^-} = Z_{nK^0}\approx 0.0090\\ \nonumber
Z_{nK^+} & = & \frac{1}{2}\,Z_A\frac{r}{r+1} +Z_{pK^-} = Z_{pK^0}\approx 0.00425
\end{eqnarray}
for production of $K^+ = (u\bar{s})$ and $K^0 = (d\bar{s})$.  
The factor $1/2$ in the second line of Eq.~\ref{eq:ZK} comes from quark
counting: there are two ways to choose the leading valence di-quark
for the processes in the first line, but only one way for those in the
second line.  The factor $r$ is the ratio of production of one charge
state of a $\Sigma$ hyperon to production of a $\Lambda$.  For example,
to produce a $K^+$ with an incident neutron in association with a hyperon,
the simplest process is $n\rightarrow K^+ +\Sigma^-$, whereas both
$\Lambda$ and $\Sigma^0$ contribute to production of $K^+$ by a proton.  Production
of $\Sigma$ is suppressed relative to $\Lambda$ at least by the squared mass ratio
$r=0.88 = (\Sigma/\Lambda)^2$, which is the value use for the plots.  Numerical results
are not very sensitive to this assumption.   As an example on the low side, 
we can use the ratio of $\Sigma^0/\Lambda$ measured in $e^+\,e^-\rightarrow {\rm hadrons}$, which will be an underestimate because
of the absence of the incident baryon.  If we take $r=0.25$~\cite{Beringer:1900zz} then 
the flux of $K^+$ decreases by $2$\% and the flux of neutral kaons by $8$\%.
The ratio of $K_S$ to $K_L$ remains unchanged.

For production of
anti-kaons we take
\begin{equation}
Z_{p\bar{K}^0}=Z_{n\bar{K}^0}=Z_{pK^-}=Z_{nK^-}\approx 0.0028.
\label{eq:ZbarK}
\end{equation}
The approximate numerical values shown here apply at energies below which
the knee in the primary spectrum affects the integral in Eq.~\ref{eq:TIG-Z}.
For energies $E_\nu > 300$~TeV the magnitude of the Z-factors decrease by
approximately a common factor as a result of the steepening of the primary spectrum. 

Next we combine the proton and neutron contributions into a single factor
that can be multiplied by the total flux of nucleons as in Eq.~\ref{eq:angular}.  So, for example, $Z_{NK^+} = f_p Z_{pK^+} + f_n Z_{nK^+}$, etc.  We take the fraction of protons as $f_p\approx 0.8$ and the neutron fraction as $f_n\approx 0.2$, appropriate for the model of the primary spectrum we are using~\cite{Gaisser:2012zz}.
With the numerical values in Eqs.~\ref{eq:ZK},\ref{eq:ZbarK}, this gives $Z_{NK^+}\approx 0.00805$.  We therefore estimate
\begin{equation}
\label{eq:Kch2nue}
\frac{\nu_e}{(\nu_e+\bar{\nu}_e)}=\frac{Z_{NK^+}}{Z_{NK^+}+Z_{NK^-}}\approx 0.74
\end{equation}
for neutrinos from decay of charged kaons.

$K_L$ and $K_S$ are orthogonal mixtures of $K^0$ and $\bar{K}^0$ with equal weights.
Therefore
\begin{equation}
Z_{NK_L} = Z_{NK_S} = \frac{1}{2}(Z_{NK^0}+Z_{N\bar{K}^0})\approx 0.0040.
\label{eq:ZKL}
\end{equation}
From Eq.~\ref{eq:ZK},   $Z_{NK^0} = f_n Z_{nK^0}+f_p Z_{pK^0}\approx 0.0052$ and
$Z_{N\bar{K}^0}\approx 0.0028$.  Thus, in the high energy limit where the Ke3 decays
reflect the $K^0$-$\bar{K}^0$ asymmetry, the ratio $\nu_e/(\nu_e+\bar{\nu}_e)\approx 0.65$ for neutrinos from decay of neutral kaons.  The composite $\nu_e$ fraction is shown
in Fig.~\ref{fig:oscillation}.

\end{document}